\documentclass[reprint,
superscriptaddress,
amsmath,amssymb,
aps,
prb,
]{revtex4-2}

\usepackage{graphicx}
\usepackage{dcolumn}
\usepackage{bm}
\usepackage{hyperref}

\begin{document}


\title{Ultrafast Relaxation Dynamics of Spin-Density Wave Order\\ in BaFe$_2$As$_2$ under High Pressures}

\author{Ivan Fotev}
\affiliation{Helmholtz-Zentrum Dresden-Rossendorf, 01328 Dresden, Germany}
\affiliation{Technische Universit\"at Dresden, 01062 Dresden, Germany}

\author{Stephan Winnerl}
\affiliation{Helmholtz-Zentrum Dresden-Rossendorf, 01328 Dresden, Germany}

\author{Saicharan Aswartham}
\affiliation{Leibniz IFW Dresden, 01069 Dresden, Germany}

\author{Sabine Wurmehl}
\affiliation{Leibniz IFW Dresden, 01069 Dresden, Germany}
\affiliation{Institute of Solid State and Materials Physics and W\"urzburg-Dresden Cluster of Excellence ct.qmat, Technische Universit\"at Dresden, 01062 Dresden, Germany}

\author{Bernd B\"uchner}
\affiliation{Leibniz IFW Dresden, 01069 Dresden, Germany}
\affiliation{Institute of Solid State and Materials Physics and W\"urzburg-Dresden Cluster of Excellence ct.qmat, Technische Universit\"at Dresden, 01062 Dresden, Germany}

\author{Harald Schneider}%
\affiliation{Helmholtz-Zentrum Dresden-Rossendorf, 01328 Dresden, Germany}

\author{Manfred Helm}%
\affiliation{Helmholtz-Zentrum Dresden-Rossendorf, 01328 Dresden, Germany}
\affiliation{Technische Universit\"at Dresden, 01062 Dresden, Germany}

\author{Alexej Pashkin}%
\email{a.pashkin@hzdr.de}
\affiliation{Helmholtz-Zentrum Dresden-Rossendorf, 01328 Dresden, Germany}

\date{\today}

\begin{abstract}
BaFe$_2$As$_2$ is the parent compound for a family of iron-based high-temperature superconductors as well as a prototypical example of the spin-density wave (SDW) system. In this study, we perform an optical pump-probe study of this compound to systematically investigate the SDW order across the pressure-temperature phase diagram. The suppression of the SDW order by pressure manifests itself by the increase of relaxation time together with the decrease of the pump-probe signal and the pump energy necessary for complete vaporization of the SDW condensate. We have found that the pressure-driven suppression of the SDW order at low temperature occurs gradually in contrast to the thermally-induced SDW  transition. Our results suggest that the pressure-driven quantum phase transition in BaFe$_2$As$_2$ (and probably other iron pnictides) is continuous and it is caused by the gradual worsening of the Fermi-surface nesting conditions.
\end{abstract}

\maketitle


\section{\label{sec:intro}Introduction}

Iron pnictide high-$T_c$ superconductors have become one of the most intensively studied material systems since their discovery in 2008 \cite{Kamihara2008}. In contrast to superconducting cuprates, they demonstrate a high sensitivity to the applied stress, for instance, to the hydrostatic pressure. The magnetic spin-density wave (SDW) order in a parental compound for these materials can be suppressed not only by chemical doping, but also by pressure \cite{Kimber2009}. Moreover, since parental iron pnictides are metallic even in the absence of doping, they can undergo pressure-induced SDW-to-superconductor quantum phase transition (QPT) at low temperatures \cite{Colombier2009,Nakai2010}. Exploring the mechanism of this transition, the competition and coexistence of the magnetic order and the superconducting condensate is crucial for the understanding of high-$T_c$ superconductivity in iron pnictides.  

\begin{figure}
	\includegraphics[scale=0.55]{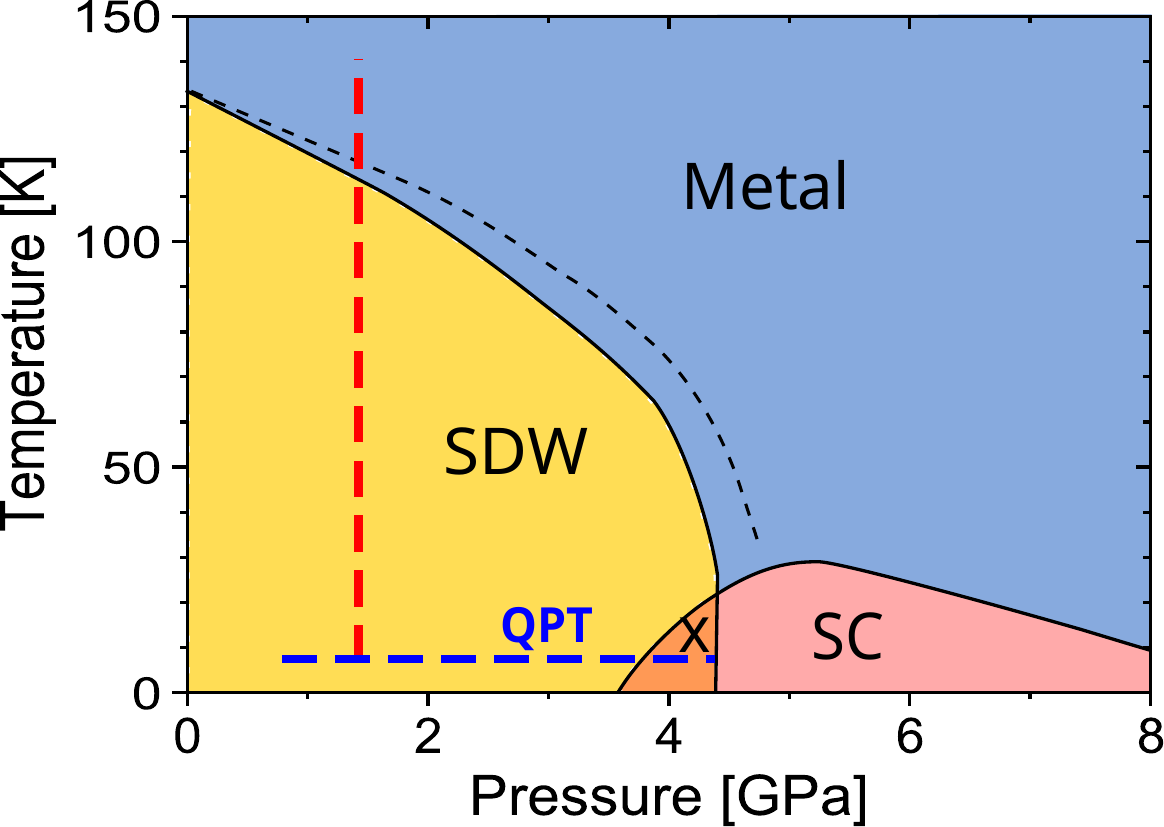}
	\caption{\label{fig:pT_scans}Schematic pressure-temperature phase diagram of BaFe$_2$As$_2$, mostly according to \cite{Colombier2009}. The black dashed line denotes the structural phase transition that splits out of the SDW transition with doping \cite{Chu09} or pressure \cite{Wu2013}. Broken red and blue lines illustrate $p=\text{const}$ and $T=\text{const}$ measurement trajectories corresponding to measurements shown in Figs.\ref{fig:DeltaR_T} and \ref{fig:QPT}, respectively.}
\end{figure}

For the present study we have chosen BaFe$_2$As$_2$ as a prime example of a well-known parental iron pnictide compound, in which the electronic state can be controlled by external pressure. At ambient pressure the onset of the magnetic SDW order occurs simultaneously with the structural transition from tetragonal to orthorhombic phase at $T_{\text{SDW}}$=137~K \cite{Aswartham2012}. A chemical doping or an external pressure application cause a decrease of the transition temperature $T_{\text{SDW}}$ \cite{Chu09}. The $p$-$T$ phase diagram of BaFe$_2$As$_2$ depicted in Fig.~\ref{fig:pT_scans} is qualitatively similar to the one of chemically doped compounds: at low temperatures the system is in the SDW state that is gradually suppressed by increasing the doping level or pressure until the superconductivity (SC) sets in \cite{Paglione2010}. 
Upon applying high pressures, the structural properties of BaFe$_2$As$_2$ change the same way as in the chemically doped materials \cite{Kimber2009}. The structural changes are correlated with the reduction of the Fermi-surface nesting between the inner hole and electron pockets, which results in the SDW state suppression above a pressure of $\approx$3\,GPa (depending on the experimental conditions) \cite{Kimber2009,Alireza2009,Colombier2009}. This suppression results in the quantum critical behavior \cite{Kasahara2010} indicating that the pressure- or doping-driven SDW phase transition should be of the second-order in contrast to the first-order temperature-induced SDW transition \cite{Sachdev2011}.

Measuring the pressure dependence of the SDW or superconducting energy gaps poses a significant challenge, because it has to be performed on a sample inside a diamond anvil cell (DAC). This precludes the use of angle-resolved photoemission spectroscopy (ARPES), which requires a direct access to the sample's surface in vacuum. Tunneling spectroscopy under high pressures was demonstrated for conventional superconductors \cite{Zhu2015}. However, it has not been applied to high-$T_c$ iron-based superconductors yet. Nevertheless, a recent tunneling spectroscopy study demonstrates a strong impact of the local surface strain on the superconducting gap in LiFeAs \cite{Cao2021}. 
Up to now, the SDW and the superconducting energy gaps in BaFe$_2$As$_2$ under high pressure have been determined using infrared spectroscopy \cite{Uykur2015}. That study reveals the coexistence of the SDW and SC orders at 3.6\,GPa since the spectral weight of the SDW gap excitation is not significantly affected by the emergence of the superconductivity.

Besides infrared spectroscopy that probes the linear optical response of a material, nonlinear pump-probe spectroscopy provides another way to estimate the energy gaps in high-$T_c$ superconductors via the analysis of quasiparticle relaxation dynamics \cite{Demsar2020}. The main advantage of pump-probe spectroscopy is its high sensitivity to the evolution of electronic order in close vicinity of phase transitions. Numerous pump-probe spectroscopy studies of iron pnictides have demonstrated that the suppression of the SDW or SC order around the transition temperatures is accompanied by a critical slowing down of the relaxation dynamics for the given electronic order \cite{Chia2010,Stojchevska2010,Mertelj2010}. Near-infrared pump – mid-infrared probe spectroscopy of parental BaFe$_2$As$_2$ has shown that the coherent lattice oscillation that modulates the Fe-As-Fe bonding angle can periodically induce transient SDW ordering, even at temperatures above $T_{\text{SDW}}$ \cite{Kim2012}.

Here we report a study of the SDW order in BaFe$_2$As$_2$ across the $p-T$ phase diagram. Our results demonstrate a critical slowing down of the SDW relaxation time caused by the suppression of SDW order both by temperature or pressure. However, in contrast to the abrupt change across the temperature-driven first-order transition, the increase of the relaxation time caused by pressure at 8~K is gradual, indicating the second-order character of the quantum phase transition.

\section{Experimental Results}

\subsection{Sample preparation and experimental setup}

The studied BaFe$_2$As$_2$ monocrystals were grown by a self-flux high temperature solution growth technique \cite{Aswartham2011, Aswartham2012}. The sample, 100$\,\mu$m $\times$ 85$\,\mu$m in size and a thickness of about 30$\,\mu$m, was initially cleaved from the BaFe$_2$As$_2$ monocrystal, so that it can fit into the sample chamber of the DAC. The commercial closed-cycle cryostat system with integrated DAC (\textit{Diacell}$^{\textrm{®}}$ \textit{CCS-DAC} from \textit{Almax}$\cdot$\textit{easyLab}) enabled us to perform \textit{in situ} control of pressure and temperature independently due to the gas membrane-driven DAC system. A Ti:sapphire amplifier system operating at the repetition rate of 250\,kHz generates 60\,fs long laser pulses utilized in the experiment. The pump and the probe beams (both with the wavelength of 800\,nm) are focused onto the sample inside the DAC, and the reflected probe signal is collected using a confocal microscopy scheme. A ruby chip was placed inside the sample chamber, next to the sample, and used for external pressure calibration \cite{Jayaraman1983}. We chose CsI powder as the pressure-transmitting medium in order to ensure a direct contact between the sample and the diamond anvil.

\subsection{Temperature dependence at fixed pressure}

\begin{figure}
\includegraphics[width=1\columnwidth]{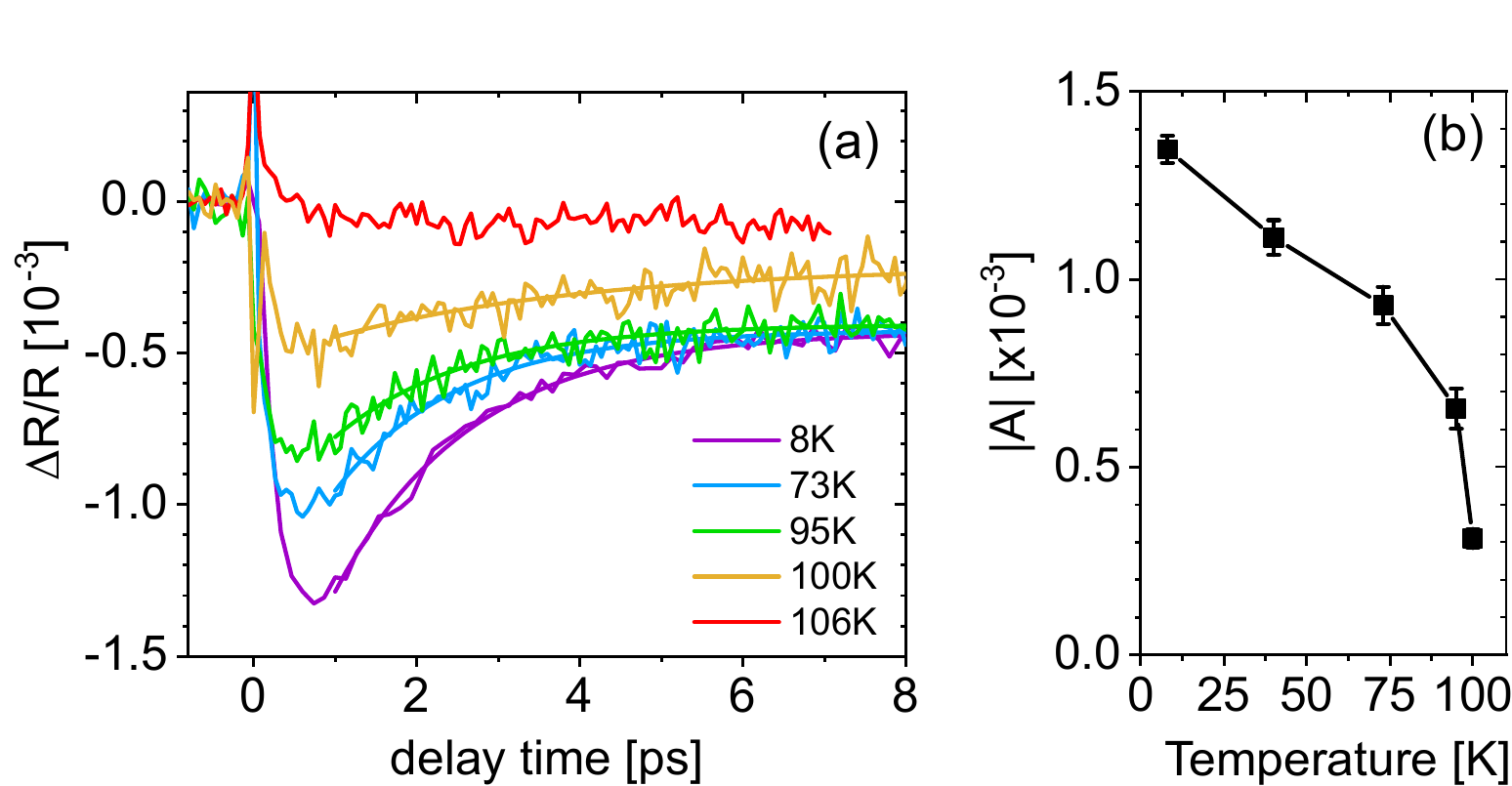}
\caption{\label{fig:DeltaR_T}(a) Relative differential reflectivity transients and their monoexponential fits, recorded at $p=1.4$\,GPa for several temperatures, and fixed pump fluence of $75\,\mu$J/cm$^2$. (b) Amplitudes (black) and relaxation times (blue) from the monoexponential fits. The substantial drop in amplitude above 100\,K, accompanied by the increase in relaxation times, reveals the phase transition from SDW to metallic state.}
\end{figure}

Figure~\ref{fig:DeltaR_T}(a) shows a set of photoinduced differential reflectivity $\Delta R(t)/R$ measurements, obtained at pump fluence of $F=75\,\mu$J/cm$^2$, and fixed pressure of 1.4\,GPa.
Together these measurement points correspond to the vertical path in the $p$-$T$ diagram, marked by the broken red line in Fig.~\ref{fig:pT_scans}.
At lower temperatures BaFe$_2$As$_2$ is in the SDW phase, and the photoinduced reflectivity change $\Delta R/R$ reaches its negative maximum quickly after the photoexcitation at the zero delay time. Then, the signal decays within few picoseconds to a nearly constant level corresponding to a thermalized hot state. We assign this monoexponential, fast-decaying part of the signal to the relaxation of the SDW phase. Its amplitude $|A|$ is proportional to the concentration of photoexcited quasiparticles: $|A|\propto n_{pe}$ \cite{Demsar1999, Gedik2004, Torchinsky2010}.
With increasing temperature $A$ decreases substantially, and near 100\,K the relaxation process slows down. Finally, the pump-probe response at $T=106$\,K (red line) does not contain the SDW relaxation signature anymore and demonstrates only a small signal with  opposite sign with a much faster decay. We assign it to the hot-electron cooling process typical for metals.

Figure~\ref{fig:DeltaR_T}(b) shows the temperature dependence of the exponential decay amplitude. Its strong drop around 100\,K is related to the SDW phase transition to the non-magnetic metallic phase. Naturally, this transition temperature determined for the pressure of 1.4\,GPa is lower than $T_{\text{SDW}} = 137$\,K at atmospheric pressure. Simultaneously, we observe indications of the critical slowing down of the relaxation dynamics that was previously reported for SmFeAsO \cite{Mertelj2010}, SrFe$_2$As$_2$ \cite{Stojchevska2010} and BaFe$_2$As$_2$ \cite{Chia2010, Stojchevska2012}. However, it was not possible to reliably quantify this effect since close to $T_{\text{SDW}}$, where the slowing down is expected, the signal amplitude becomes very small and the fitting uncertainty is too large.

At temperatures well below $T_{\text{SDW}}$ the relaxation time $\tau \approx 1.6$\,ps remains nearly constant. This behavior contradicts the bimolecular recombination model: $\mathrm{d}n_{pe}/\mathrm{d}t = -2R n_T n_{pe} -R n_{pe}^2$, where $R$ is the bimolecular recombination coefficient and $n_T$ and $n_{pe}$ are the densities of thermally and photoexcited quasiparticles, respectively. Here the first term $R n_T n_{pe}$ corresponds to the recombination of a photoexcited quasiparticle with a thermally populated one, and the second term $R n_{pe}^2$ represents the recombination of the two photoexcited quasiparticles \cite{Torchinsky2011}. At low temperatures the bimolecular model predicts that the initial recombination rate $\tau^{-1}$ should increase with $n_T$, which strongly depends on temperature. Similar to our observations, the studies of \text{SmFeAsO} \cite{Mertelj2010} and \text{SrFe$_2$As$_2$} \cite{Stojchevska2010} also report $\tau$ that does not increase with decreasing temperatures. It is interpreted as a hint for an additional relaxation channel besides the bimolecular recombination at the SDW gap; namely, an interband scattering of quasiparticles to other, ungapped electronic bands \cite{Stojchevska2010}. Such scattering should be temperature independent and it should result in an additional, dominating monomolecular recombination term $-n_{pe}/\tau$ in the rate equation.

\subsection{Suppression of the SDW relaxation by pressure and temperature}

\begin{figure}
\includegraphics[width=0.85\columnwidth]{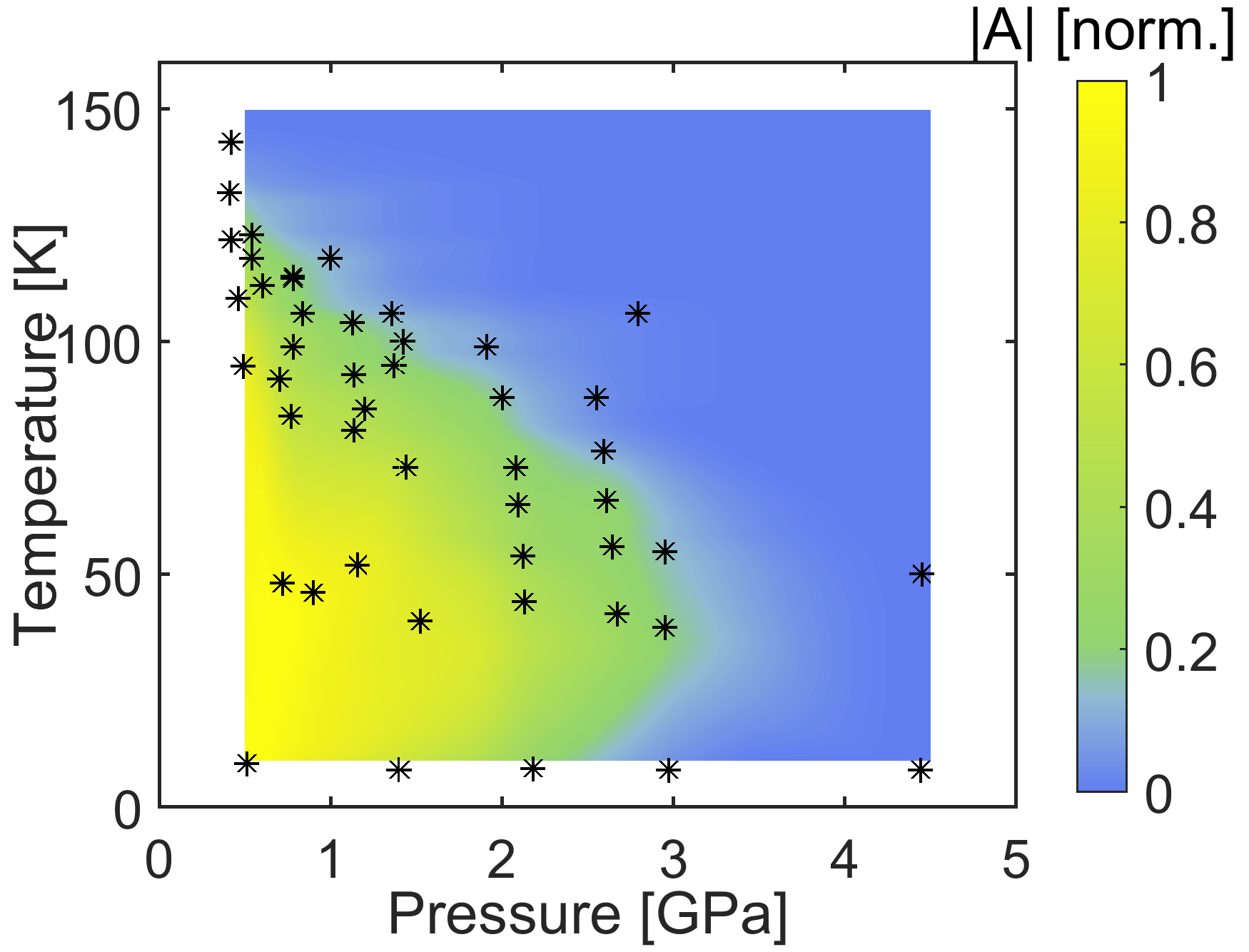}
\caption{\label{fig:Amplitudes}$p-T$ diagram obtained from the amplitudes of the SDW transients $A$. At the phase transitions to metallic or superconducting states the amplitude goes to zero (blue color). The individual measurement points are represented by asterisks.}
\end{figure}

The data from all temperature-dependent measurements at fixed pressures from 0.5 to 4.4\,GPa (vertical scans of the $p$-$T$ diagram) are combined in a $p-T$ diagram, depicted in Fig.~\ref{fig:Amplitudes}. The color represents the amplitude of the normalized monoexponential signal $|A|$: from yellow for 1, to blue for 0.
Therefore, the metallic phase corresponds to the blue region of the diagram and the density of the SDW condensate is displayed by the brightness of the yellow color.
As expected from Fig.~\ref{fig:pT_scans}, the temperature at which the amplitude vanishes, i.e. the SDW transition temperature $T_{\text{SDW}}(p)$, decreases with pressure, and at high pressures the SDW state gets completely suppressed.
From the 4.4\,GPa measurements we could not identify SDW $\Delta R(t)/R$ transients anymore, so we assigned zeros to the amplitude of those points. 

As shown in Figs.~\ref{fig:DeltaR_T}(b) and \ref{fig:Amplitudes}, upon increasing temperature the amplitudes approach smoothly and monotonically zero at the corresponding $T_{\text{SDW}}(p)$ for each pressure, which was also reported in Ref.~\onlinecite{Chia2010}. This is in contrast to the other two studies of iron pnictides \cite{Stojchevska2010, Mertelj2010}, where the temperature dependence of $A(T)$ exhibits a small increase relatively close to $T_{\text{SDW}}$, before starting to drop rapidly: a characteristic feature of the bottleneck model \cite{Kabanov1999}. However, the pump fluences used in the present study are much larger than in the other three reports, making the direct comparison harder.

\subsection{Relaxation dynamics and its fluence dependence at 8~K}

\begin{figure}
\includegraphics[width=0.8\columnwidth]{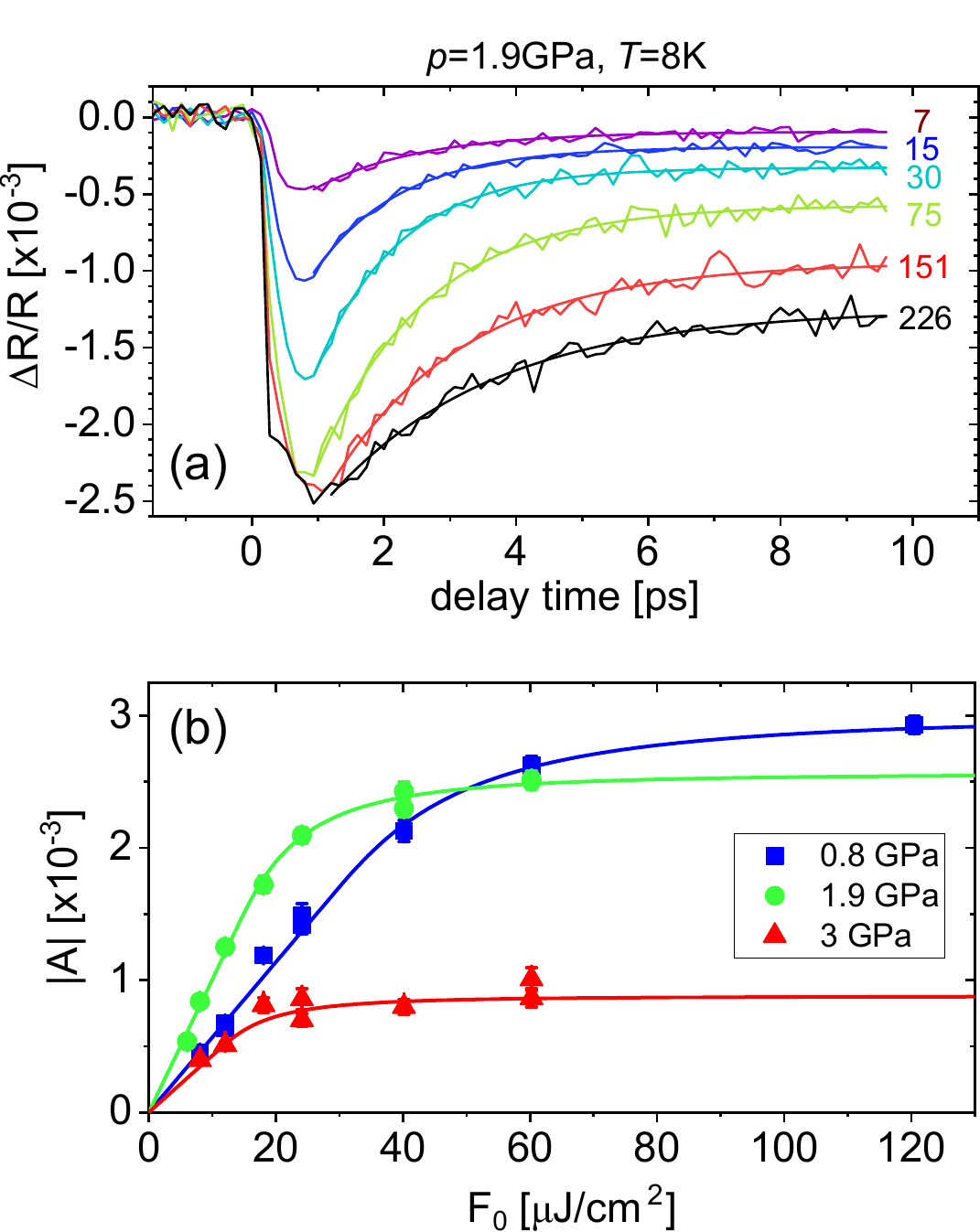}
\caption{\label{fig:Saturation}(a) Reflectivity changes $\Delta R(t)/R$ measured at 1.9\,GPa and 8\,K for selected pump fluences. The fluence values are given in $\mu$J/cm$^2$. (b) Fluence dependence of the SDW reflectivity transient amplitudes $|A|$, for several pressures. $F_0$ is the part of the pump fluence $F$ which penetrates into the sample, after the reflection from its surface. Lines are the fits with the model of the threshold fluence for condensate vaporization from \cite{Kusar2008}.}
\label{fig:Saturation}
\end{figure}

Now let us examine the suppression of the SDW order by pressure in the ground state, i.e., the path of the quantum phase transition as it is depicted in Fig.~\ref{fig:pT_scans}. In order to gain more reliable information about the SDW phase we measured pump-probe response for various pump fluences at each pressure in the range from 0.8 to 3\,GPa. Figure~\ref{fig:Saturation}(a) presents a set of measurements taken at different pump fluences at pressure of 1.9\,GPa and temperature of 8\,K.
Starting from $7\,\mu$J/cm$^2$ the amplitude $A$ grows linearly with pump fluence $F$, roughly up to $30\,\mu$J/cm$^2$, after which it completely saturates.
This happens because the proportionality between the number of the photoexcited quasiparticles and the number of the pump photons $n_{pe} \propto F$, which is valid in the low-fluence regime, breaks down when the pump fluence becomes high enough for a noticeable depletion of the SDW condensate. Finally, at a certain threshold vaporization fluence $F_T$ the SDW condensate is completely suppressed, and consequentially no more quasiparticles can be excited. As a result, $n_{pe}$ and the pump-probe signal $A$ get saturated.

\begin{figure}
	\centering
	\includegraphics[width=0.75\columnwidth]{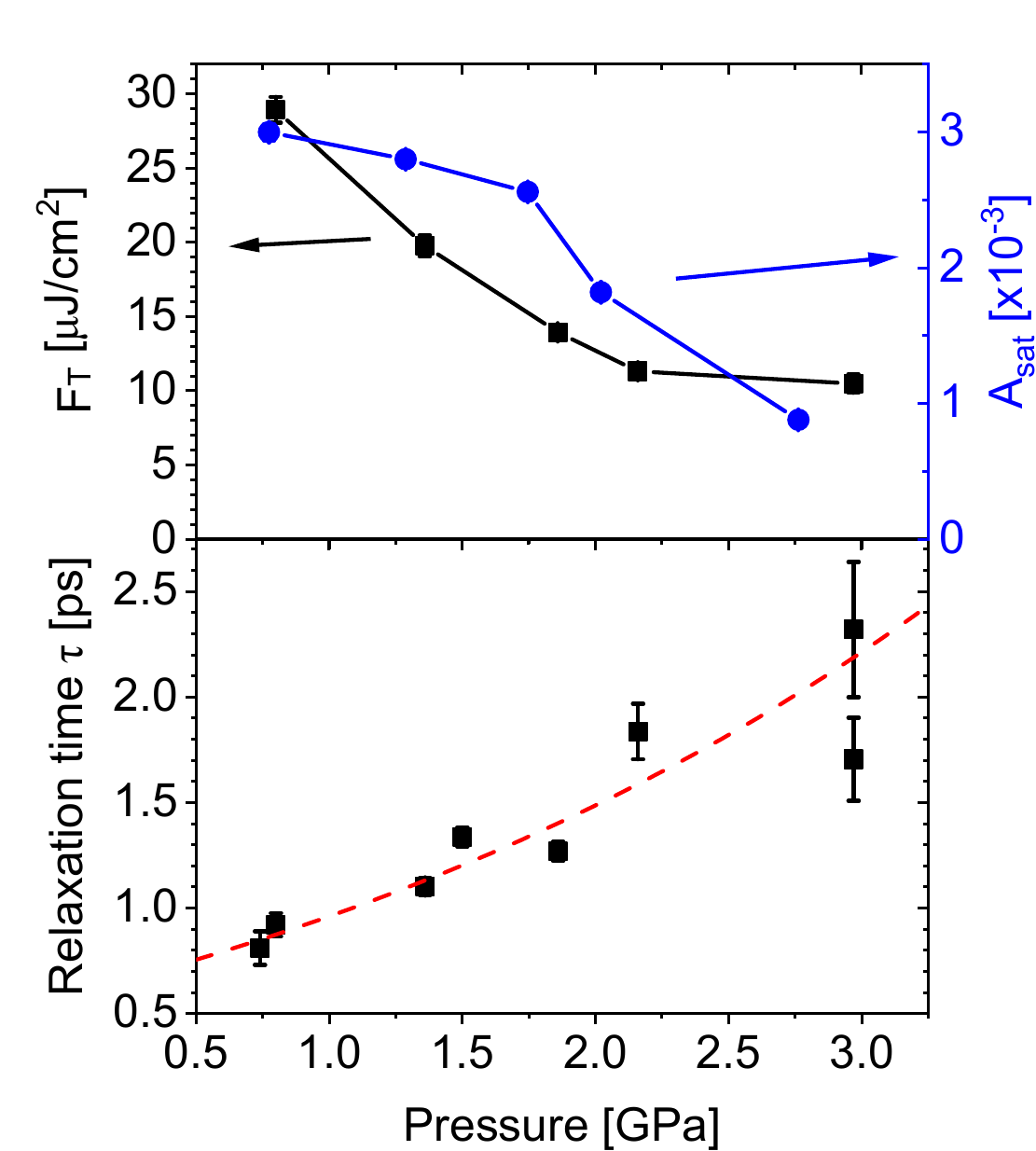}
	\caption{(a) Pressure dependence of the threshold vaporization fluence $F_T$ (black squares) and saturation amplitudes (blue circles). The value of $F_T$ can be considered as the measure of SDW condensation energy. (b) Pressure dependence of the relaxation time $\tau(p)$ at fixed temperature $T=8\,K$ and fluence of $30\,\mu$J/cm$^2$.}
	\label{fig:QPT}
\end{figure}

The saturation curves measured at a few chosen pressures are shown in Fig.~\ref{fig:Saturation}(b). Here $F_0 = (1-R)F$ is the part of the pump fluence $F$ which penetrates the sample, after the reflection from its surface. $R = 0.2$ is the reflectivity of BaFe$_2$As$_2$ (at 800\,nm) with respect to the diamond that is almost independent on pressure and temperature \cite{Uykur2015}. Although the curves look qualitatively similar, the saturation fluence $F_T$ as well as the saturation amplitude $A_{sat}$ decrease with pressure. We have extracted both parameters from the fits based on the model introduced in Ref.~\onlinecite{Kusar2008}, which takes into account the lateral Gaussian and exponential depth profiles of the pump and probe beams. The fitting curves are represented by lines in Fig.~\ref{fig:Saturation}(b). The obtained values of $F_T(p)$ and $A_{sat}(p)$ for all measured pressures are presented in Fig.~\ref{fig:QPT}(a) with black squares and blue circles, respectively.

Besides the suppression of the pump-probe signal and the decrease of the saturation fluence, the applied pressure also affects the relaxation time $\tau$. Figure~\ref{fig:QPT}(b) shows its pressure dependence $\tau(p)$, for a fixed $T$=8\,K and $F$=30\,$\mu$J/cm$^2$. Similarly to the temperature-dependent measurements \cite{Chia2010, Stojchevska2012} the increasing $\tau$ signalizes the vicinity of the phase transition. As the system approaches the transition pressure, the gap $\Delta_{\text{SDW}}(p)$ gets smaller, and consequently the SDW recovery time slows down. It is known that near the transition point the relaxation time is inversely proportional to the gap \text{$\tau \propto \Delta^{-1}$} \cite{Schmid1975, Schuller1976, Demsar2020}.  Thus, all these observations indicate the gradual suppression of the SDW state by the external pressure leading to the quantum phase transition in BaFe$_2$As$_2$. Since at 4.4\,GPa we do not observe any SDW relaxation signature (see Appendix~\ref{suppl}), the pressure of the QPT should be between 3 and 4.4~GPa.

\section{Discussion}

Similar to the case of the superconducting condensate in cuprates \cite{Kusar2008} it is useful to compare the energy deposited at the saturation fluence with the thermodynamical condensation energy that can be obtained from the specific heat measurements \cite{Rotter2008}. For the lowest pressure measurement at $p = 0.8\,$GPa the threshold fluence is $F_T=29\,\mu$J/cm$^2$ (Fig.~\ref{fig:QPT}(a)). In order to estimate the absorbed energy density we use the optical penetration depth of BaFe$_2$As$_2$ at $1.55$\,eV (800\,nm) to be $\lambda_{op} = 36\,$nm. This calculation is based on the reported real parts of dielectric constant $\varepsilon_1 \approx 1$ and conductivity $\sigma_1 = 1500\,\Omega^{-1}$cm$^{-1}$ taken from Ref.~\onlinecite{Barisic2010}. Strictly speaking, the mentioned values of $\varepsilon_1$ and $\sigma_1$ were measured on the optimally doped compound, but since the reflectivity $R$ is the same at 1.55\,eV for the doped and parent compounds \cite{Lucarelli2010}, we assume that this is also holds for $\varepsilon_1$ and $\sigma_1$, and consequently, $\lambda_{op}$. Therefore, the absorbed energy density at the vaporization threshold is $F_T/\lambda_{op}=8\,$J/cm$^3$. On the other hand, the thermodynamic SDW condensation energy is estimated to be $E_c/V = 1.6\,$J/cm$^3$, using the specific heat $C(T)$ data for the parent BaFe$_2$As$_2$ ($p=0\,$GPa) \cite{Rotter2008}. Thus, the pump energy necessary for the vaporization of the SDW condensate is about 5 times larger than the condensation energy density of BaFe$_2$As$_2$. A similar result was reported for La$_{2-x}$Sr$_x$CuO$_4$ cuprate superconductors, and interpreted as the evidence that a large portion of the pump energy goes to the phonon subsystem and does not affect the electronic condensate \cite{Kusar2008}. Therefore, the saturation amplitude $A_{sat}$ that is related to the full depletion of the SDW condensate appear to be more reliable parameters for the estimation of the SDW gap energy.

\begin{figure}
	\centering
	\includegraphics[width=0.7\columnwidth]{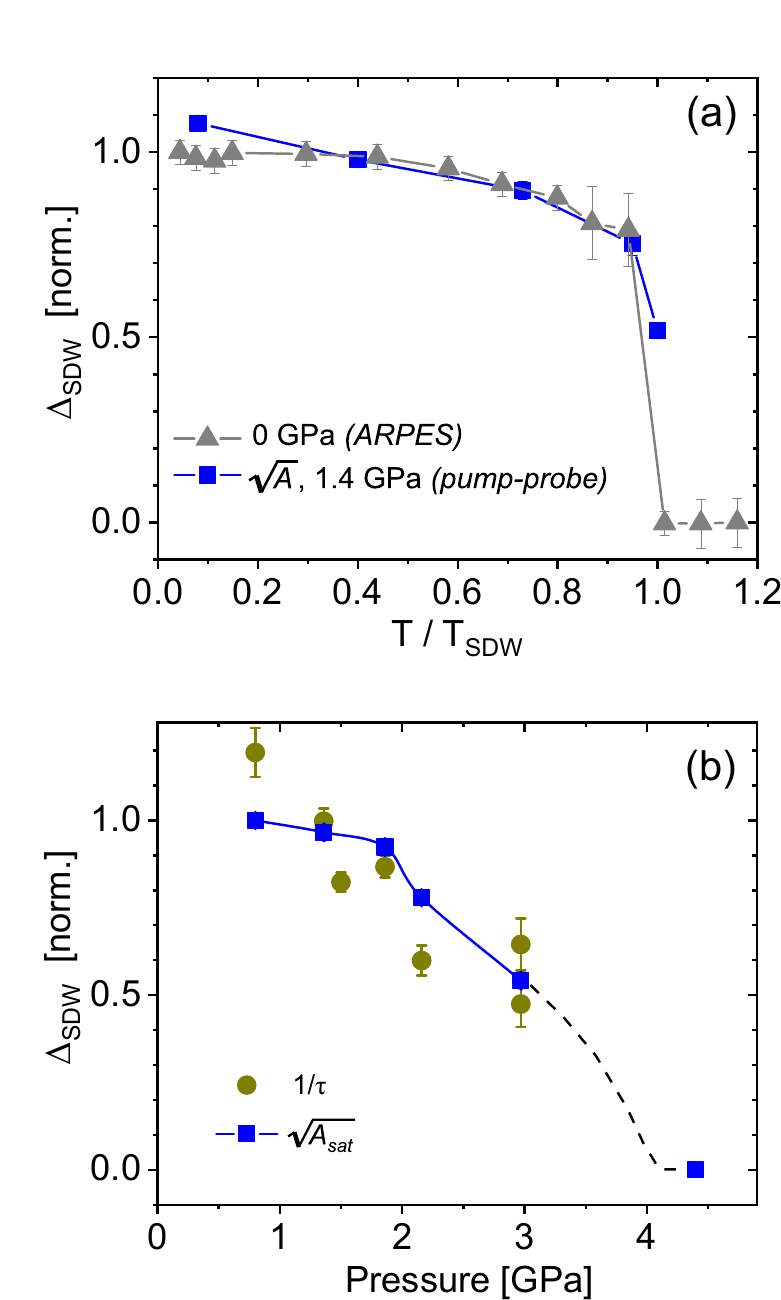}
	\caption{(a) Temperature dependence of $\sqrt{A(T)}$ (blue squares) from the measurements at $1.4\,$GPa and $F = 75\,\mu$J/cm$^2$, compared with the BaFe$_2$As$_2$ SDW gap $\Delta_{\text{SDW}}(T)$ (gray triangles), extracted from the ARPES study at $p=0\,$GPa \cite{Yi2014}. (b) The evolution of the SDW gap with pressure at $T = 10\,$K estimated as $\sqrt{A_{sat}}$ and $1/\tau$. The highest pressure point shows the absence of the SDW order at 4.4\,GPa. The dashed line schematically depicts the evolution of $\Delta_{\text{SDW}}(T)$ across the QPT.}
	\label{fig:Delta}
\end{figure}

According to the Ginzburg-Landau theory the order parameter scales as $|\Delta|^2 \propto n_{cond} \propto A_{sat}$ ($n_{cond}$ is the condensate density) and $|\Delta| \propto 1/\tau$, when $T \rightarrow T_{\text{SDW}}$.
Figure~\ref{fig:Delta}(a) shows the normalized $\sqrt{A(T)}$ dependence at $1.4\,$GPa and $F=75\,\mu$J/cm$^2$ (blue squares). Based on the results in Fig.~\ref{fig:QPT}(a), this fluence value corresponds to the saturation regime and, thus, we expect that $|\Delta_{\text{SDW}}| \propto \sqrt{A}$. For comparison, Figure~\ref{fig:Delta}(a) also shows the normalized SDW gap $\Delta_{\text{SDW}}(T)$ (gray triangles), extracted from the angle-resolved photoemission spectroscopy (ARPES) measurements \cite{Yi2014}. Since in contrast to the ARPES study, the depicted pump-probe results were recorded under high pressure ($p=1.4\,$GPa), the corresponding SDW transition temperatures are different: for our data $T_{\text{SDW}}\approx$\,100\,K and for the ARPES study $T_{\text{SDW}}=$\,138\,K. In order to compare both data sets, the horizontal axis in Fig.~\ref{fig:Delta}(a) is normalized to $T_{\text{SDW}}$. Both match quite well demonstrating that the relation $|\Delta_{\text{SDW}}| \propto \sqrt{A_{sat}}$ is fulfilled in broad range of temperatures below the SDW transition. Some deviation is observed far from the phase transition  ($T<0.3\,T_{\text{SDW}}$) where the Ginzburg-Landau theory is no longer fully applicable.

Now we apply the same method to evaluate the SDW gap as a function of pressure. Figure~\ref{fig:Delta}(b) shows the pressure dependence of $\Delta_{\text{SDW}}$ at $T$=8\,K estimated as $\sqrt{A_{sat}(p)}$ using the saturation amplitude values from Fig.~\ref{fig:QPT}(a). In addition, the evaluated relaxation time $\tau(p)$ for $F=30\,\mu$J/cm$^2$ (Fig.~\ref{fig:QPT}(b)) are reliable enough to use them for another estimation of the SDW gap as $\Delta_{\text{SDW}} \propto 1/\tau(p)$, which are depicted in Figure~\ref{fig:Delta}(b) as dark yellow circles. In general, all data in Fig.~\ref{fig:Delta}(b) indicate the trend of a \emph{gradual} decrease of $\Delta_{\text{SDW}}$ by about 50\% on the pressure increase up to 3\,GPa. This corresponds to $p / p_{\text{SDW}} > 0.7$ taking into account that $p_{\text{SDW}} < 4.4\,$GPa according to our results. If we compare Figs.~\ref{fig:Delta}(a) and (b), it is apparent that the phase transition achieved by increasing pressure happens much more gradually compared to the first-order thermally-driven phase transition. Possibly, this indicates the second-order character of the pressure-induced QPT in BaFe$_2$As$_2$ that we schematically depicted by the dashed line in Fig.~\ref{fig:Delta}(b). This conclusion agrees with the observation of quantum fluctuations near the quantum critical point in isovalently doped BaFe$_2$(As$_x$P$_{1-x}$)$_2$ \cite{Kasahara2010}. Microscopically the gradual character of the pressure-induced changes can be understood as the result of a gradual failure of the Fermi-surface nesting that causes the suppression of the SDW order \cite{Kimber2009}.

Finally, we discuss the possible pressure-induced superconducting phase. Our data demonstrate that for pressures well above 3~GPa the SDW order is fully suppressed and, according to the phase diagram (Fig.~\ref{fig:pT_scans}), one expects the onset of the superconductivity. In order to verify that, we have measured pump-probe traces $\Delta R(t)/R$ at 8\,K and 4.4\,GPa (see Appendix~\ref{suppl}). All pump-probe taken at various pump fluences show the same relaxation dynamics, which is much faster than in the SDW state. The absence of fluence-dependent relaxation dynamics and saturation behavior leads us to the conclusion that the fast dynamics observed at 4.4~GPa does not come from the superconducting or SDW state, but rather from the cooling of the hot electrons in the metallic phase.

\section{Conclusion}

We have performed a systematic temperature- and pressure-dependent pump-probe study in order to investigate the evolution of the SDW order in BaFe$_2$As$_2$. Measuring the amplitude of the relaxation process has enabled us to construct the $p-T$ phase diagram that depicts the variation of the SDW order across the whole range of the applied pressures and temperatures. The comparison of the saturation fluence with the thermodynamic condensation energy of the SDW state shows that the absorbed pump energy required for the suppression of the magnetic order largely exceeds the condensation energy indicating that a large portion of the absorbed pump energy is transferred to the phonon bath. Using the parameters of the SDW relaxation process we have estimated the variation of the SDW gap as a function of temperature and pressure. Our results demonstrate an apparent difference in the character of the thermally-induced phase transition and the pressure-driven quantum phase transition. The former is of the first-order and occurs rather abruptly near $T_{\text{SDW}}$, whereas the latter occurs gradually and the SDW gap starts to decrease for pressures well below the critical pressure $p_{\text{SDW}}$. This behavior suggests a second-order character of the quantum phase transition in BaFe$_2$As$_2$ in agreement with previous studies of quantum fluctuations in this material \cite{Kasahara2010}.

\begin{acknowledgments}
We would like to thank Jure Demsar and Ece Uykur for fruitful discussions.
This work was supported by the Deutsche Forschungsgemeinschaft DFG through the project PA 2113/1-2 and the Priority Programme SPP1458. 
\end{acknowledgments}

\appendix

\section{Pump-probe signal at 4.4\,GPa}\label{suppl}

\begin{figure*}[t]
	\centering
	\includegraphics[width=1.35\columnwidth]{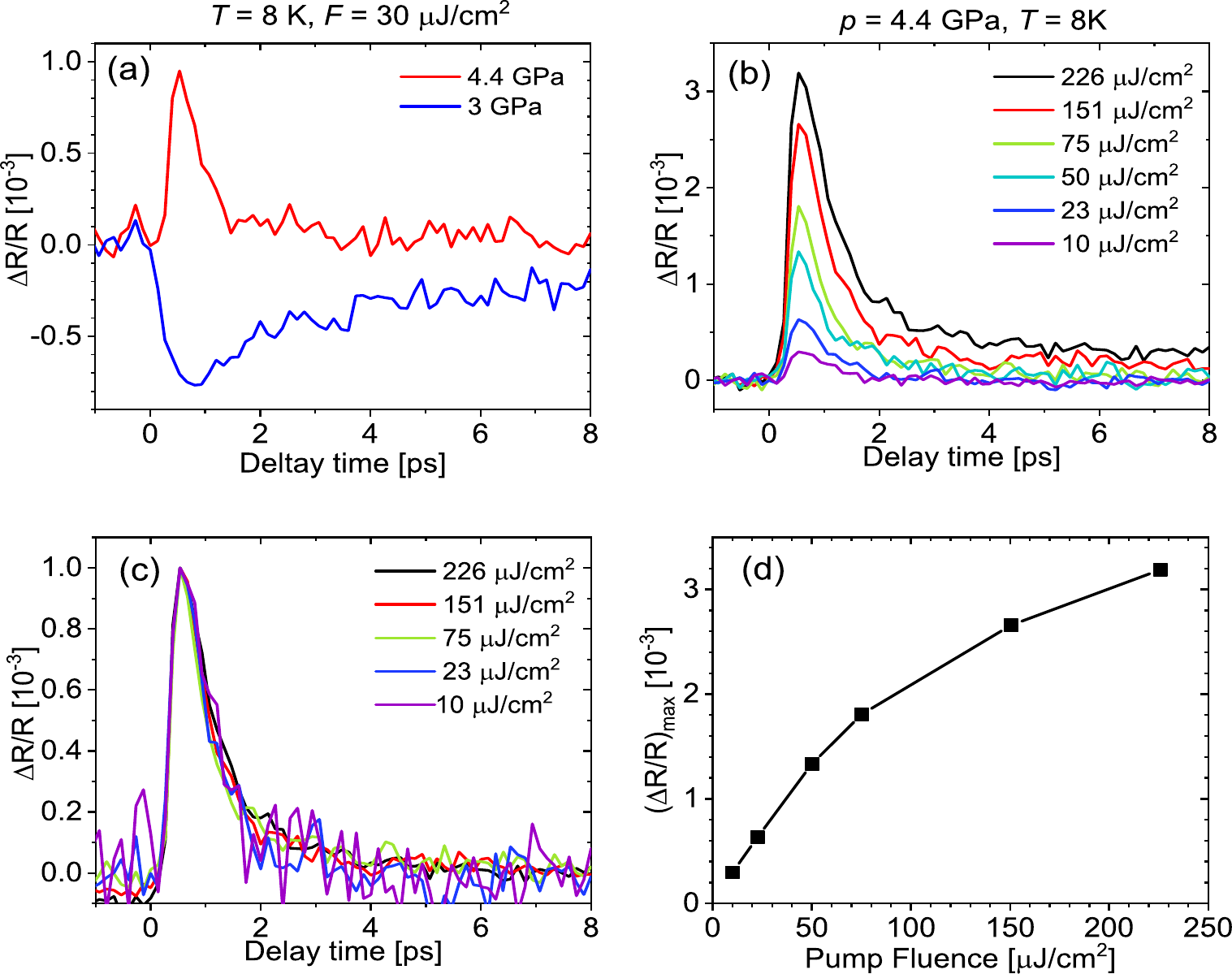}
	\caption {(a) Comparison between the reflectivity transients taken at 3\,GPa (blue), and 4.4\,GPa (red), the conditions which should correspond to the SDW and superconductivity phases, respectively. (b) Fluence dependence of the $\Delta R(t)/R$ transients at 4.4\,GPa. (c) The same data after the offsets were subtracted, and transients normalized. The lifetimes are around 0.7\,ps, and do not depend on the pump fluence. (d) Maxima of the  transients (shown in (b)) grow linearly up to $F = 75\,\mu$J/cm$^2$, and after that sub-linearly, but without saturating (at least not in the investigated fluence range up to $226\,\mu$J/cm$^2$). This indicates that the reflectivity transients might come from the electrons in the metallic phase.}
	\label{fig:DeltaR_hp}
\end{figure*}

Figure \ref{fig:DeltaR_hp}(a) shows the change in reflectivity $\Delta R(t)/R$ measured at 8\,K and 4.4\,GPa (red line).
According to the $p$-$T$ diagram, BaFe$_2$As$_2$ is expected to be in the superconducting phase at that pressure.
For the sake of comparison, we have plotted also the SDW relaxation transient at 3\,GPa (blue line), which was recorded at the same fluence and temperature.
Already the sign flip and much shorter lifetime of the 4.4\,GPa measurement suggest that the SDW phase has been indeed suppressed, i.e. the phase transition occurred.
The fluence dependence of $\Delta R(t)/R$ transients, recorded at 8\,K and 4.4\,GPa, are shown in Fig.~\ref{fig:DeltaR_hp}(b). 
The same data, normalized after subtracting the offsets, are presented in Fig.~\ref{fig:DeltaR_hp}(c). Fig.~\ref{fig:DeltaR_hp}(d) shows maxima of the relfectivity transients from Fig.~\ref{fig:DeltaR_hp}(b).

In the fluence range of $F=10–226\,\mu$J/cm$^2$, the initial relaxation time is practically constant: $\tau \approx 0.7\,$ps, and, although growing linearly up to $75\,\mu$J/cm$^2$, and sub-linearly at higher fluences, the amplitudes did not saturate.
On the contrary, the characteristic feature of many superconductors is the decrease of $\tau$ with $F$, which can be explained using the bimolecular recombination model.
Torchinsky et al. \cite{Torchinsky2010} found for the optimally hole-doped Ba$_{0.6}$K$_{0.4}$Fe$_2$As$_2$ that at 7\,K and the fluences in the range of 1.4–37\,$\mu$J/cm$^2$, the relaxation time decreases by about two orders of magnitude with increasing fluence. Furthermore, they reported that this initial, fast relaxation component is always followed by a slow relaxation process caused by a boson bottleneck, lasting for more than 100\,ps \cite{Torchinsky2010}, which we also did not observe in our high-pressure experiment.
For the optimally doped Ba$_{0.6}$K$_{0.4}$Fe$_2$As$_2$, the saturation fluence of the superconducting state was estimated to be already at $25\,\mu$J/cm$^2$, at 7\,K \cite{Torchinsky2011}.
The absence of fluence-dependent relaxation dynamics and saturation behavior (in the fluence range characteristic for the SDW state) leads us to the conclusion that the observed relaxation process is not related to the superconducting or SDW state, but perhaps rather stems from the relaxation of the hot electrons in the metallic phase.

\providecommand{\noopsort}[1]{}\providecommand{\singleletter}[1]{#1}%

\end{document}